\documentclass[aps,prb,twocolumn,superscriptaddress,amsmath,amsfonts,showpacs]{revtex4}

\usepackage{graphicx}
\usepackage{bm} 

\begin{document}


\title{ 
Topological Hall Effect 
in Inhomogeneous Superconductors
}


\author{Satoshi Fujimoto}
\affiliation{Department of Physics, Kyoto University, Kyoto 606-8502, Japan}




\date{\today}

\begin{abstract}
We propose a possible mechanism of topological Hall effect in inhomogeneous superconducting states.
In our scenario, the Berry phase effect associated with spatially modulated superconducting order parameter
gives rise to a fictitious Lorentz force acting on quasiparticles. 
In the case of the Fulde-Ferrell-Larkin-Ovchinnikov state, 
the topological Hall effect is detected by applying an electromagnetic wave with a tuned wave number 
on a surface of the system.
\end{abstract}

\pacs{}


\maketitle



In a spin-singlet superconductor under an applied
magnetic field, when the Pauli depairing effect dominates over the orbital depairing effect,
spatially modulated superconducting order parameter
is stabilized.\cite{LO,FF}
It has been discussed recently that this inhomogeneous superconducting state called
the Fulde-Ferrell-Larkin-Ovchinnikov (FFLO) state
may be realized in
a heavy fermion system CeCoIn$_5$ and some quasi-low-dimensional superconductors.\cite{FFLO,CeCoIn1,CeCoIn2, CeCoIn3,CeCoIn4,gloos,houz,gruen,adachi,Agterberg}
The possible realization of an analogous inhomogeneous superconducting state was also proposed for noncentrosymmetric superconductors, in which anti-symmetric spin-orbit interaction combined with the Zeeman magnetic field
stabilizes the helical vortex state.\cite{helical}
It is an important issue to establish the realization of 
these exotic superconducting states in the above-mentioned systems experimentally.
From this perspective, 
it is useful to study electromagnetic properties specific to the inhomogeneous superconducting states
in detail, which may be utilized for the experimental identification of the modulated order parameter.\cite{Rad,samokhin}
In this paper, we demonstrate that a spatially-varying superconducting order parameter characterizing the
inhomogeneous state gives rise to distinct electromagnetic response caused 
by topological Berry phase effects.\cite{niu}
In particular, under a certain circumstance, the topological Hall effect can be raised by
a fictitious "Lorentz force" which is generated by the Berry phase effect associated with
the inhomogeneous order parameter.
It was discussed by Bruno et al. that for electrons interacting with spin textures which possess
a nonzero Berry curvature, the Hall effect is induced by the fictitious Lorentz force raised by
the Berry phase effect.\cite{bruno} 
We here consider a possible analogous phenomenon in superconducting states with a spatially slowly-varying
order parameter.
We note that the topological Hall effect considered in this paper is 
a transport property of quasiparticles, and we do not consider the Hall 
effect associated with supercurrents here.\cite{kita}
 
Our approach is based upon
the quasiclassical method for the description of quasiparticle dynamics in
superconducting states.\cite{Eil,Larkin,vekhter}
We extend the quasiclassical Eilenberger equation to take into account important Berry phase effects.
The basic quantity with which we are concerned in the following argument is   
the single-particle Green function for the superconducting state, from which
dynamical properties can be derived;
\begin{eqnarray}
\hat{G}(x_1,x_2)=
\left(
\begin{array}{cc}
G(x_1,x_2) & -F(x_1,x_2) \\
F^{\dagger}(x_1,x_2) & \bar{G}(x_1,x_2)
\end{array}
\right)
\end{eqnarray}
where $G(x_1,x_2)$ and $F(x_1,x_2)$ are, respectively, the normal and anomalous 
Green functions, and
$\bar{G}(x,x')=G(x',x)$, $x_1=(\bm{r}_1,t_1)$ etc.
The spatial modulation of the superconducting order parameter is expressed in terms of 
the center of mass coordinate $\bm{R}=\frac{\bm{r}_1+\bm{r}_2}{2}$.
Fourier transforming the relative coordinate $\bm{r}=\bm{r}_1-\bm{r}_2$,
we introduce the Wigner transformation of the Green function,
\begin{eqnarray}
{\hat{G}}(\bm{k},\bm{R},\varepsilon_n)=\int d\bm{r}\int^{\beta}_0 d\tau{\hat{G}}(\bm{R}+\frac{\bm{r}}{2},\frac{\tau}{2},\bm{R}-\frac{\bm{r}}{2},-\frac{\tau}{2})&&
\nonumber \\
\times e^{i\varepsilon_n\tau}e^{i\bm{k}\bm{r}},&&
\label{foug}
\end{eqnarray}
with $\varepsilon_n$ the fermionic Matsubara frequency.
When there is a vector potential $\bm{A}_0$,
the Gor'kov equation satisfied by ${\hat{G}}(\bm{k},\bm{R},\varepsilon_n)$ is
\begin{eqnarray}
\left[{\hat{\tau}}_3i\varepsilon_n-\varepsilon(\bm{k}-\frac{i}{2}\nabla_{\bm{R}}
-e\bm{A}_0\hat{\tau}_3
)+{\hat{\tau}}_3h-{\hat{\Delta}}
-{\hat{\Sigma}}\right] && \nonumber \\
\times{\hat{G}}(\bm{k},\bm{R},\varepsilon_n)={\hat{1}}&&
\label{gorkov}
\end{eqnarray}
where $\varepsilon(\bm{k})$ is the energy band dispersion for electrons, $h=\mu_{\rm B}H_z$ is the Zeeman magnetic field, 
${\hat{\tau}}_\mu$ ($\mu=1,2,3$) is the Pauli matrix in the particle-hole space,
and $\hat{\Sigma}$ is the normal selfenergy matrix which is diagonal in the particle-hole space, and includes effects of impurity scattering,
and electron-electron interaction.
We consider the spin-singlet pairing state with the gap function,
\begin{eqnarray}
{\hat{\Delta}}(x,x')=
 \left(
\begin{array}{cc}
0 & \Delta(\bm{R}) \\
-\Delta^{*}(\bm{R}) & 0
\end{array}
\right).
\end{eqnarray}
We expand the kinetic energy term of (\ref{gorkov}) in terms of the spatial gradient $\nabla_{\bm{R}}$, and transform the basis of the particle hole space as
$\tilde{\hat{G}}=\hat{G}\hat{\tau}_3$. Then, eq.(\ref{gorkov}) is rewritten into
\begin{eqnarray}
\left[i\varepsilon_n+{\hat{\tau}}_3\frac{i}{2}\bm{v}\nabla_{\bm{R}}+e\bm{v}\bm{A}_0
+h-{\hat{\bm{\tau}}}\cdot\hat{\bm{H}}_0
-{\hat{\tau}}_3{\hat{\Sigma}}\right] && \nonumber \\
\times\tilde{\hat{G}}(\bm{k},\bm{R},\varepsilon_n) ={\hat{1}} &&
\label{gorkov2}
\end{eqnarray}
with $\hat{\bm{H}}_0=(\Delta_1(\bm{R}),-\Delta_2(\bm{R}),\varepsilon(\bm{k}))$,
$\Delta_1({\bm{R}})={\rm Re}\Delta(\bm{R})$, $\Delta_2(\bm{R})={\rm Im}\Delta(\bm{R})$, and
$\bm{v}=\frac{\partial \varepsilon(\bm{k})}{\partial \bm{k}}$.
We diagonalize the fifth term $\hat{\bm{\tau}}\cdot\hat{\bm{H}}_0$ by applying the unitary transformation
$\tilde{\hat{G}}'=\hat{U}^{\dagger}(\bm{R})\tilde{\hat{G}}\hat{U}(\bm{R})$; i.e.
$\hat{U}^{\dagger}(\bm{R})\hat{\bm{\tau}}\cdot\hat{\bm{H}}_0\hat{U}(\bm{R})=E(\bm{k})\hat{\tau_3}$
 and $E(\bm{k})=\sqrt{\varepsilon^2(\bm{k})+|\Delta(\bm{R})|^2}$:
\begin{eqnarray}
\bigl[i\varepsilon_n+\frac{i}{2}\hat{U}^{\dagger}{\hat{\tau}}_3\hat{U}\bm{v}\nabla_{\bm{R}}
+\frac{\bm{v}}{2}\hat{\bm{A}}_{\rm f}
 +e\bm{v}\bm{A}_0+h-E(\bm{k})\hat{\tau}_3  &&\nonumber \\
-\hat{U}^{\dagger}{\hat{\tau}}_3{\hat{\Sigma}}\hat{U}\bigr] 
\tilde{\hat{G}}' ={\hat{1}}. &&
\label{gorkov3}
\end{eqnarray}
Here, the unitary transformation applied to the spatial gradient term of (\ref{gorkov2}),
${\hat{\tau}}_3\frac{i}{2}\bm{v}\nabla_{\bm{R}}$,
gives rise to a fictitious vector potential 
$\hat{\bm{A}}_{\rm f}=i\hat{U}^{\dagger}\hat{\tau}_3\nabla_{\bm{R}}\hat{U}$, which is a $2\times2$ matrix in the particle-hole space.
At this stage, we apply the adiabatic approximation, assuming that
the transition between the electron band with the energy $E(k)$ and the hole band with the energy $-E(k)$
is suppressed, and
neglect the off-diagonal terms of $\hat{U}^{\dagger}\hat{\tau}_3\hat{U}$ and $\hat{\bm{A}}_{\rm f}$; 
i.e. 
\begin{eqnarray}
\hat{U}^{\dagger}\hat{\tau}_3\hat{U}\rightarrow \frac{\varepsilon(k)}{E(k)}\hat{\tau}_3,
\label{utu}
\end{eqnarray}
\begin{eqnarray}
 \hat{\bm{A}}_{\rm f}\rightarrow \frac{\varepsilon(k)}{E(k)}(\bm{A}_{\rm f1}+\bm{A}_{\rm f2}\hat{\tau}_3),
 \label{ficvecapp}
\end{eqnarray} 
with 
$\bm{A}_{{\rm f} 1}=\frac{1}{2}(1-\frac{E(k)}{\varepsilon(k)})\nabla_{\bm{R}}\phi(\bm{R})$, and
$\bm{A}_{{\rm f} 2}=-\frac{i}{4}\frac{\nabla_{\bm{R}}|\Delta|^2}{E^2(k)}$.
Here $\phi(\bm{R})$ is the phase of the gap function; $\Delta(\bm{R})=|\Delta(\bm{R})|e^{i\phi(\bm{R})}$.
The approximation (\ref{utu}) and (\ref{ficvecapp}) is most crucial in our argument for the realization of the Berry phase effect. The Berry phase effect appears when one restricts the Hilbert space within a sub-space in which the change of the phase of the wave function is regarded as an adiabatic one.  Here, we restrict the Hilbert space within the electron band or the hole band.
Then, suppressing the transition between the electron band and the hole band,
one can neglect the off-diagonal elements of Eq.(\ref{gorkov3}), as done in Eqs.(\ref{utu}) and (\ref{ficvecapp}). This approximation is valid at sufficiently low temperatures, because of the energy difference between the electron band and the hole band due to the Zeeman splitting. 
We will discuss the validity of the approximation in more detail in the last part of this paper.
As a result, the diagonal component (\ref{ficvecapp}) can be regarded as
fictitious U(1) gauge fields acting on quasiparticles.
Within this approximation, $\tilde{\hat{G}}'$ is diagonal;  $\tilde{\hat{G}}'={\rm diag}(\tilde{G}'_{+},\tilde{G}'_{-})$.
Each component satisfies,
\begin{eqnarray} 
\biggl[i\varepsilon_n-\sigma E(\bm{k}-\frac{i}{2}\nabla_{\bm{R}}-\sigma\frac{\bm{A}_{\rm f1}}{2}-\frac{\bm{A}_{\rm f2}}{2}-\sigma e\frac{E(k)}{\varepsilon(k)}\bm{A}_0) && \nonumber \\
+h-\tilde{\Sigma}_{\sigma}'\biggr]
\tilde{G}'_{\sigma}=1, &&
\label{gorkov-dia}
\end{eqnarray}
with $\sigma=\pm$, and $\tilde{\Sigma}_{\sigma}'$ is the diagonal component of $\hat{U}^{\dagger}\hat{\tau}_3
\hat{\Sigma}\hat{U}$.
In Eq.(\ref{gorkov-dia}), we have rewritten the derivative term and the vector potential terms into a
gauge invariant form.

To simplify the analysis, we assume that $\bm{A}_0$ is a time-dependent uniform field, which yields an electric field.
It is straightforward to generalize the following analysis to the case that $\bm{A}_0$ also produces a magnetic field.
To solve (\ref{gorkov-dia}) for $\tilde{G}'_{\sigma}$, we follow the quasiclassical approach developed by Eilenberger.
We extract the left-hand Gor'kov equation from
the right-hand equation (\ref{gorkov-dia}), 
expand it in terms of the spatial gradient $\nabla_{\bm{R}}$
up to the second order, and integrate each term over the energy dispersion $\varepsilon_k\equiv \varepsilon(\bm{k})$.
From the second term of (\ref{gorkov-dia}), we obtain,
\begin{eqnarray}
\int \frac{d\varepsilon_k}{\pi}\left[i\sigma\frac{\varepsilon_k}{E(k)}\bm{v}\tilde{\nabla}_{\bm{R}}\tilde{G}'_{\sigma}+
i(\bm{v}\times \bm{B}_{\rm f})\frac{\partial \tilde{G}'_{\sigma}}{\partial \bm{k}_{\parallel}}
\right],
\label{ficlor}
\end{eqnarray}
where $\bm{B}_{\rm f}$ is the Berry curvature $\bm{B}_{\rm f}=\nabla\times \bm{A}_{\rm f1}$, the explicit expression of which is
\begin{eqnarray}
(\bm{B}_{\rm f})_{\alpha}=-\epsilon_{\alpha\beta\gamma}\frac{1}{4E^2(k)}\frac{\partial |\Delta|^2}{\partial R_{\beta}}
\left(\frac{\partial \phi}{\partial R_{\gamma}}-2eA_{0\gamma}\right)
\label{ficb}
\end{eqnarray}
and $\tilde{\nabla}_{\bm{R}}$ is the derivative with respect to $\bm{R}$ under the constraint that
$E(k-\sigma\frac{\bm{A}_{\rm f1}}{2}-\frac{\bm{A}_{\rm f2}}{2}-\sigma e \frac{E_k}{\varepsilon_k}\bm{A}_0)$ is fixed.
$\bm{k}_{\parallel}$ is the momentum parallel to the Fermi surface.
Note that the $\bm{A}_{\rm f2}$ term of (\ref{gorkov-dia}) is a pure gauge, and does not give a nonzero Berry curvature.
The second term of Eq.(\ref{ficlor}) is the fictitious "Lorentz force" term, the origin of which is 
the topological Berry phase effect raised by 
the spatial modulation of the superconducting order parameter. 
From (\ref{ficb}),
we see that the fictitious magnetic field is nonzero only
when both the amplitude and the phase of the superconducting gap
are spatially modulated.
Thus, the topological Hall effect does not occur for the Fulde-Ferrel state and the helical vortex phase, in which only the phase of the superconducting gap is modulated.\cite{FF,helical}
It is also noted that the fictitious magnetic field (\ref{ficb}) has a gauge-invariant form.

In the standard quasiclassical approach, the Gor'kov equation is recast into the Eilenberger equation for
the normalized Green function,
\begin{eqnarray}
\tilde{g}'_{\sigma}=\int \frac{d\varepsilon_k}{\pi}{\tilde{G}'_{\sigma}}.
\end{eqnarray}
However, unfortunately, the first term of (\ref{ficlor}) can not be expressed in terms of the normalized Green function because of a strongly varying factor $\varepsilon_k/E(k)$, which stems from the use of the transformed Green function $\tilde{\hat{G}}'$ instead of the standard Green function $\hat{G}$.
To avoid this difficulty, we restrict our argument within the case with a uniform current, and discard 
this term.
For the second term of (\ref{ficlor}), we evaluate the integral over $\varepsilon_k$ in the following manner.
\begin{eqnarray}
\int \frac{d\varepsilon_k}{\pi}
(\bm{v}\times \bm{B}_{\rm f})\frac{\partial \tilde{G}'_{\sigma}}{\partial \bm{k}_{\parallel}}
\approx(\bm{v}\times\tilde{\bm{B}}_{\rm f})\frac{\partial \tilde{g}'_{\sigma}}{\partial \bm{k}_{\parallel}},
\label{lorentz}
\end{eqnarray}
with 
\begin{eqnarray}
\tilde{\bm{B}}_{\rm f}=\left\{
\begin{array}{ll}
\bm{B}_{\rm f}|_{\varepsilon_k=0}&
\mbox{for }  \Delta(\bm{R}) > h \\[0.2cm]
0& \mbox{for }  \Delta(\bm{R}) < h.
\end{array}
\right.
\label{br}
\end{eqnarray}
Then, the Eilenberger equation for the uniform current state satisfied by the normalized Green function
$\tilde{g}'_{\sigma}(\varepsilon_n,\varepsilon_{n'})$ under the vector potential $\bm{A}_0(t)=\bm{A}_0(\omega_0)e^{i\omega_0 t}$
is given by
\begin{eqnarray}
&& (i\varepsilon_n-i\varepsilon_{n'})\tilde{g}'_{\sigma}(\varepsilon_n,\varepsilon_{n'})
+i(\bm{v}\times\tilde{\bm{B}}_{\rm f})\cdot\nabla_{\bm{k}_{\parallel}}\tilde{g}'_{\sigma}(\varepsilon_n,\varepsilon_{n'})
\nonumber \\
&&+e\bm{v}\bm{A}_0(\omega_0)[\tilde{g}'_{\sigma}(\varepsilon_n-\omega_0,\varepsilon_{n'})
-\tilde{g}'_{\sigma}(\varepsilon_n,\varepsilon_{n'}+\omega_0)]  \nonumber \\
&&-\tilde{\sigma}_{\sigma}(\varepsilon_n)\tilde{g}'_{\sigma}(\varepsilon_n,\varepsilon_{n'})+\tilde{g}'_{\sigma}(\varepsilon_n,\varepsilon_{n'})\tilde{\sigma}_{\sigma}(\varepsilon_{n'})=0.
\label{eil1}
\end{eqnarray}
Here $\tilde{\sigma}_{\sigma}(\varepsilon_n)$ is the normalized selfenergy $\tilde{\sigma}_{\sigma}=\int \frac{d\varepsilon_k}{\pi}\tilde{\Sigma}'_{\sigma}$. We have neglected effects of external fields on the selfenergy.
The second term of (\ref{eil1}) is the fictitious Lorentz force term which gives rise to the topological Hall effect.

In the following, to be concrete, we consider the case of the FFLO state
with the spatially modulated superconducting gap $\Delta(\bm{R})=\Delta_0\cos qx$,
which is believed to be realized in CeCoIn$_5$ under an applied magnetic field parallel 
to the $x$-axis.\cite{CeCoIn1,CeCoIn2}
We examine the Hall current parallel to the $x$-axis induced by the topological Berry phase effect, 
when an electric field $E_y$ is applied along the $y$-axis.
Note that since the Hall current is parallel to the external magnetic field in this situation, its is not difficult to distinguish between the topological Hall effect considered here and the ordinary Hall effect induced by the applied magnetic field
in experimental measurements.
As mentioned above, to obtain
the nonzero $\bm{B}_{\rm f}$, we need the spatial modulation of the phase of the superconducting gap $\phi$, as well as the amplitude modulation due to the FFLO state.
To fulfill this requirement, we consider the situation that the electric field $E_y$ induces a 
supercurrent parallel to the $y$-axis, and thus
$\partial_y \phi -2eA_{0y}\neq 0$. 
We solve Eq.(\ref{eil1}) for $\tilde{g}'_{\sigma}$ in the vicinity of
the superconducting transition temperature $T_c$.
It is easily seen from (\ref{eil1}) that $\tilde{g}'_{+}$ and $\tilde{g}'_{-}$ in the uniform current state are the same.
Thus, we obtain
the normalized Green function in the original particle-hole space 
$\hat{g}={\rm diag}(g,\bar{g})=\tilde{g}'_{+}\hat{\tau}_3$.
Then, the expression for the Hall current for $T\sim T_c$ is
\begin{eqnarray}
J_x^{\rm Hall}&=&T\sum_{n}\int d\Omega_{\bm{k}}ev_x(g-\bar{g})|_{i\omega\rightarrow \omega+i\delta \atop \omega\rightarrow 0} \nonumber \\
&\approx& \frac{\sigma_n\tau\langle (\tilde{\bm{B}}_{\rm f})_z\rangle}{m}E_y.
\label{Hallc}
\end{eqnarray}
Here $\langle (\tilde{\bm{B}}_{\rm f})_z\rangle$ is the spatial average of the fictitious magnetic field, $\sigma_n$ is the normal state conductivity, and $\tau$ 
is the relaxation time of electrons.
Note that from the derivation described above, it is apparent that the topological Hall effect considered here
is a non-linear response to external fields.
The bulk Hall current obtained above is nonzero only when the spatial average of the fictitious field
$\tilde{\bm{B}}_{\rm f}$ is nonzero.
This imples that $\langle \partial_x|\Delta|\rangle=|\Delta(L_x)|-|\Delta(0)|\neq 0$; i.e.
the magnitude of the gap function at the two opposite edges must be different.
This condition crucially depends on extrinsic factors such as the geometry of 
a sample used for the measurement of the Hall effect, and pinning of the
nodal plane of the FFLO state due to impurities.
These extrinsic factors, unfortunately, makes it difficult to detect the Hall current
experimentally. 
To avoid such extrinsic factors, one can use the STM measurement for the detection of the Hall effect.
Even when the condition $\langle \partial_x|\Delta|\rangle=|\Delta(L_x)|-|\Delta(0)|\neq 0$
is not satisfied, the fictitious Lorentz force induced by the Berry curvature is balanced by
the electrostatic force due to the topological Hall voltage which has a spatial dependence
$V_{\rm Hall}\sim \cos qx$ in the above-mentioned model.
This electrostatic field gives rise to the inhomogeneous charge redistribution, which may be observed on
the surface of the system via the STM measurement.
It should be cautioned that the charge disproportion raised by the topological Hall effect can not be described by Eq.(\ref{Hallc}), because it is assumed in its derivation that the current is spatially uniform.

However, more promising approach for the detection of the topological Hall effect is to exploit 
an electromagnetic wave $\bm{E}_0e^{i(\omega t-kx)}$ applied on a surface of the system.
For the setup considered above, the electromagnetic wave is a monochromatic plane wave
propagating along the $x$-axis, and  is linearly polarized so as that
the electric field $\bm{E_0}$ is parallel to the $y$-axis. 
We consider the situation that this electromagnetic wave is applied in addition to
the static electric field parallel to the $y$-axis, which is required to
realize the nonzero fictitious field $\tilde{\bm{B}}_{\rm f} \neq 0$.
When the wave number $k$ is chosen to be equal to $q$,
the oscillating factor of the fictitious magnetic field $\tilde{\bm{B}}_{\rm f}$ is cancelled out with
that of the electromagnetic field, and hence, there is the net nonzero fictitious Lorentz force
acting on quasiparticles.
In this situation, we obtain the ac topological Hall current flowing along the $x$-direction on the surface,
which is easily detected.
Since the induced Hall current is uniform in this case, the derivation of the expression of the Hall conductivity 
presented above is justified, and the Hall current 
is given by (\ref{Hallc}) with $\langle (\tilde{\bm{B}}_{\rm f})_z\rangle$ replaced with
$\langle (\tilde{\bm{B}}_{\rm f})_z\cos qx\rangle$.
It is noted that the above argument can be straightforwardly extended to the case with the orbital effect of 
magnetic fields.\cite{gruen,adachi}
We stress again that the direction of the topological Hall current considered here is parallel to
the applied external magnetic field, which is required to realize the FFLO state.
Thus, in experimental measurements, one can clearly discriminate between the topological Hall effect and the ordinary Hall effect of quasiparticles induced by the applied magnetic field.\cite{note}

Finally, we discuss the validity of the adiabatic approximation which is crucial in our argument.
The emergence of the Berry phase effect is due to
the application of the adiabatic approximation; i.e. the transition between
the electron band and the hole band is neglected, which is the central assumption
in the derivation of the Gor'kov equation with the fictitious vector potential (\ref{gorkov-dia}).
This assumption is valid as far as there is an energy gap which separates
the electron band and hole band, and temperature is sufficiently lower than the energy scale of the gap.
However, in the FFLO state with the gap function $\Delta(x)=\Delta_0\cos qx$, 
there are nodal planes of the superconducting gap at which $\Delta(x)=0$.
The existence of the nodal planes affects the energy spectrum of quasiparticles drastically.
This issue was solved exactly in the case of the one-dimensional system,\cite{Machida,samokhin}
and it was found that there is still an energy gap in the quasiparticle spectrum, which may
validate the adiabatic approximation.
However, in two and three dimensions, with which we are concerned,
it may be possible that the quasiparticle spectrum may become gapless,
because of the energy dispersion in the direction perpendicular to the $x$-axis.
Nevertheless, we can justify the adiabatic approximation applied to our system because of
the following reason.
Even if the superconducting gap vanishes at the nodal plane of the FFLO state,
there is still an energy gap between the electron band with up (down) spin 
and the hole band with down (up) spin in the vicinity of the Fermi level, 
because of the Zeeman splitting.
Thus, for temperatures much lower than the Zeeman energy scale, the transition between
these two bands is suppressed, and hence,
the adiabatic approximation is properly applied.
Although we derived the expression for the topological Hall conductivity (\ref{Hallc})
only in the vicinity of the transition temperatures, the Hall effect is more clearly observed at sufficiently low temperatures.

In summary, we have demonstrated that the topological Hall effect of quasiparticles can be raised by
the Berry phase effect associated with the spatially slowly-varying superconducting order parameter 
which is realized in the FFLO state.
The experimental detection of this effect which is feasible with the use of 
an ac electromagnetic field applied on a surface of the system may provide
an evidence of the realization of the inhomogeneous superconducting state.

We thank K. Samokhin for illuminating discussions, and introducing the author ref.\cite{Machida}.
This work is supported by the Grant-in-Aids for
Scientific Research from MEXT of Japan
(Grants No.19052003 and No.21102510).


\begin{thebibliography}{99}
\bibitem{note} Although the oscillating magnetic field induced by the applied electromagnetic wave is 
parallel to the $z$-axis, it does not raise the net Hall current along the $x$-axis, and not affect our result for
the topological Hall effect.
\end{thebibliography}
\end{document}